\PassOptionsToPackage{table}{xcolor}
\documentclass[sigconf,screen]{acmart} 

\AtBeginDocument{%
  }

\usepackage{makecell}
\usepackage{tabularx}
\usepackage{booktabs}
\usepackage{tabularray}
\usepackage{threeparttable}

\usepackage{subcaption}

\usepackage{balance}

\newcommand{\customsize}{\fontsize{8.5}{10.5}\selectfont}
\newenvironment{smalltabularx}{\customsize \tabularx}{\endtabularx}
\newcommand{\customtinysize}{\fontsize{8}{10}\selectfont}

\newcolumntype{C}{>{\centering\arraybackslash\hsize=.5\hsize\linewidth=\hsize}X}

\usepackage{enumitem}
\setlist[itemize]{leftmargin=*,itemsep=-0.1em,topsep=0.2em, partopsep=0.2em}

\usepackage{tcolorbox}
\usepackage{soul}
\usepackage{pifont}
\newcommand{\cmark}{\ding{51}}%
\newcommand{\xmark}{\ding{55}}%

\usepackage{threeparttable}
\usepackage{multirow}

\usepackage[nameinlink,noabbrev]{cleveref}

\usepackage{siunitx} 
\usepackage{fp}
\newcommand{\perc}[1]{%
\footnotesize
  \FPifneg{#1}
    \textcolor{orange!60!black}{\footnotesize $\downarrow$\num[round-mode=places,round-precision=1]{\stripminus #1}\%}
  \else
    \textcolor{green!40!black}{\footnotesize \textbf{$\uparrow$\num[round-mode=places,round-precision=1]{#1}\%}}
  \fi
}

\newcommand{\percplus}[1]{%
  \FPifneg{#1}%
    \textcolor{orange!60!black}{\footnotesize$\downarrow$\num[round-mode=places,round-precision=2]{\stripminus #1}\%}%
  \else%
    \textcolor{green!40!black}{\footnotesize\textbf{$\uparrow$\num[round-mode=places,round-precision=2]{#1}\%}}%
  \fi%
}

\newcommand{\stripminus}[1]{\expandafter\the\numexpr-#1}

\copyrightyear{2025}
\acmYear{2025}
\setcopyright{cc}
\setcctype{by}
\acmConference[ISSTA Companion '25]{34th ACM SIGSOFT International Symposium on Software Testing and Analysis}{June 25--28, 2025}{Trondheim, Norway}
\acmBooktitle{34th ACM SIGSOFT International Symposium on Software Testing and Analysis (ISSTA Companion '25), June 25--28, 2025, Trondheim, Norway}\acmDOI{10.1145/3713081.3731745}
\acmISBN{979-8-4007-1474-0/2025/06}

\newcommand{\framework}{\textsc{Cama}}

\begin{document}

\title{On Benchmarking Code LLMs for Android Malware Analysis} 



\author{Yiling He}
\affiliation{%
  \institution{University College London}
  \city{London}
  \country{United Kingdom}
  }
\email{yiling-he@ucl.ac.uk}

\author{Hongyu She}
\affiliation{%
  \institution{Zhejiang University}
  \city{Hangzhou}
  \country{China}
}
\email{hongyushe@zju.edu.cn}

\author{Xingzhi Qian}
\affiliation{%
  \institution{University College London}
  \city{London}
  \country{United Kingdom}
  }
\email{xingzhi.qian.23@ucl.ac.uk}

\author{Xinran Zheng}
\affiliation{%
  \institution{University College London}
  \city{London}
  \country{United Kingdom}
  }
\email{xinran.zheng.23@ucl.ac.uk}

\author{Zhuo Chen}
\affiliation{%
  \institution{Zhejiang University}
  \city{Hangzhou}
  \country{China}
}
\email{hypothesiser.hypo@zju.edu.cn}

\author{Zhan Qin}
\affiliation{%
  \institution{Zhejiang University}
  \city{Hangzhou}
  \country{China}
}
\email{qinzhan@zju.edu.cn}

\author{Lorenzo Cavallaro}
\affiliation{%
  \institution{University College London}
  \city{London}
  \country{United Kingdom}
  }
\email{l.cavallaro@ucl.ac.uk}

\renewcommand{\shortauthors}{Yiling He, Hongyu She, et al.}

\begin{abstract}

Large Language Models (LLMs) have demonstrated strong capabilities in various code intelligence tasks.
However, their effectiveness for Android malware analysis remains underexplored. 
Decompiled Android malware code presents unique challenges for analysis, due to the malicious logic being buried within a large number of functions and the frequent lack of meaningful function names.

This paper presents \framework{}, a benchmarking framework designed to systematically evaluate the effectiveness of Code LLMs in Android malware analysis. 
\framework{} specifies structured model outputs to support key malware analysis tasks, including \textit{malicious function identification} and \textit{malware purpose summarization}. 
Built on these, it integrates three domain-specific evaluation metrics{}\textemdash{}\textit{consistency}, \textit{fidelity}, and \textit{semantic relevance}{}\textemdash{}enabling rigorous stability and effectiveness assessment and cross-model comparison.

We construct a benchmark dataset of $118$ Android malware samples from $13$ families collected in recent years, encompassing over $7.5$ million distinct functions, and use \framework{} to evaluate four popular open-source Code LLMs.
Our experiments provide insights into how Code LLMs interpret decompiled code and quantify the sensitivity to function renaming, highlighting both their potential and current limitations in malware analysis.

\end{abstract}




\maketitle

\section{Introduction}

Recent advances in Large Language Models~(LLMs) have transformed natural language processing~\cite{radford2018improving}, and their extension to code understanding has led to the emergence of Code LLMs, i.e., specialized LLMs trained on large-scale code repositories.
These models have achieved strong performance in tasks such as code generation, summarization, and repair~\cite{li2023cctest, tian2024debugbench, llmcode24usenix}.
Given their growing capabilities in reasoning about code, Code LLMs offer a promising direction for automating Android malware analysis, a domain where analysts must often manually inspect large volumes of low-level code to uncover malicious behavior~\cite{mantovani2022re}.


While the potential is clear, effectively applying Code LLMs in this context remains challenging. 
First, decompiled Android code is often obfuscated, lacks type information, and contains incomplete control structures~\cite{meng2016binary, shang2024far}, diverging sharply from the clean, structured code these models are typically trained on~\cite{zheng2023codegeex}.
Second, accurate semantic interpretation is difficult due to the high-level and diverse malicious behaviors in malware~\cite{wang2022malradar}, while the absence of reliable function-level ground truth (e.g., labeling functions as malicious or benign) further complicates evaluation~\cite{he2023finer, miller, salem}. 
These challenges underscore the need for a structured evaluation framework that systematically assesses and compares Code LLM performance in real-world malware analysis scenarios.

To systematically evaluate the performance of \ul{C}ode LLMs in \ul{A}ndroid \ul{m}alware \ul{a}nalysis, we define a structured output format comprising three key elements: \textit{function summaries}, \textit{refined function names}, and \textit{maliciousness scores}. 
While function summaries are commonly generated by Code LLMs to describe the purpose of code snippets~\cite{husain2019codesearchnet}, refined function names address the lack of meaningful identifiers in decompiled code and aid analysts in quickly understanding a function's intent. 
Additionally, maliciousness scores explicitly quantify the potential security risks associated with each function, serving as critical indicators for malicious behavior localization~\cite{downing2021deepreflect}. 
As these structured outputs constitute an interpretable and actionable representation, they offer potential to support both human analysts and automated systems in malware analysis~\cite{humanvsmachine, he2024dream}. 

We consider two key malware analysis tasks to benchmark LLM performance: \textit{malicious function identification} and \textit{malware purpose summarization}. 
For each task, we propose tailored domain-specific metrics. Specifically, for malicious function identification, we define 1)~\textit{Consistency}, measuring the stability of generated function names and maliciousness scores under a self-referential process, and 2)~\textit{Fidelity}, quantifying how effectively LLM-generated maliciousness scores distinguish between benign and malicious functions. For malware purpose summarization~\cite{qian2025lamd}, we introduce 3)~\textit{Semantic Relevance}, assessing how well aggregated function-level summaries and refined function names generated by the LLM align with ground-truth malware descriptions.

We implement our evaluation framework as \framework{} and demonstrate its applicability through a detailed case study. 
Specifically, we construct a benchmark dataset consisting of $118$ Android malware samples across $6$ categories and $13$ families, collectively comprising over $7.5$ million distinct functions. 
We select $4$ popular open-source Code LLMs~(i.e., CodeLlama~\cite{roziere2023codellama}, StarChat~\cite{li2023starcoder}, CodeT5~\cite{wang2021codet5}, and PLBART~\cite{ahmad2021unified}) and design tailored prompting and tuning strategies to generate the desired structured outputs.
This study investigates two key research questions: 1)~How well do Code LLMs interpret decompiled Android code in malware analysis? and 2)~How does function renaming influence the effectiveness of LLM-based analysis?

For the first question, we analyze the quality of LLM-generated outputs using our structured output format and domain-specific metrics. For the second, we systematically rename functions in the original decompiled code with the LLM-suggested names and measure the subsequent impact on evaluation metrics.
Our findings reveal that while Code LLMs can generate informative summaries, their understanding of maliciousness remains limited. 
Among existing Code LLMs, instruction-tuned GPT-style models significantly outperform Seq2Seq models across different metrics; function renaming enhances fidelity and consistency but may reduce semantic clarity, indicating a trade-off that warrants careful consideration.


\noindent\textbf{Contributions.}  
This work makes the following key contributions:
\begin{itemize}
    \item We propose a benchmarking framework for evaluating Code LLMs for Android malware analysis, incorporating structured outputs and downstream malware analysis tasks. We further define three domain-specific metrics—consistency, fidelity, and semantic relevance—to rigorously assess the stability and effectiveness of LLM-generated outputs.
    \item We construct a benchmark dataset of 118 representative Android malware samples and a total of 7,542,799 distinct functions to demonstrate our framework's utility. Our empirical analysis provides critical insights into Code LLMs’ capabilities in interpreting decompiled code and quantifies the impact of function renaming on malware analysis outcomes. 
\end{itemize}

\section{Background and Related Work}

\subsection{Code Large Language Models}

Code LLM refers to large language models specifically trained on programming-related data to assist with coding tasks. 
These models are pre-trained on extensive code repositories~\cite{bairi2024codeplan}, documentation~\cite{khan2021automatic}, and other technical resources~\cite{pearce2023examining}, equipping them with a strong understanding of syntax, semantics, and programming patterns. When fine-tuned with different datasets or optimization techniques, Code LLMs can be tailored to excel in specific tasks, such as code completion, translation, and summarization, across multiple programming languages~\cite{wang2023recode, chen2023teaching, khan2023xcodeeval}. These capabilities make them valuable tools for automating and streamlining various aspects of the software development process.


\begin{table}[t]
\centering
\begin{threeparttable}
\begin{smalltabularx}{.96\linewidth}{cccCC}
    \toprule
    \textbf{Model} & \textbf{Style} & \textbf{Architecture} & \textbf{Java\tnote{*}} & \textbf{Inst.\tnote{\dag}} \\
    \midrule
    CodeT5~\cite{wang2021codet5} & T5 & Encoder-Decoder & \cmark & \xmark \\
    PLBART~\cite{ahmad2021unified} & BART & Encoder-Decoder & \cmark & \xmark \\
    CodeLlama~\cite{roziere2023codellama} & GPT & Decoder-only & \cmark & \cmark \\ 
    StarChat~\cite{li2023starcoder} & GPT & Decoder-only & \cmark & \cmark \\ 
    \bottomrule
\end{smalltabularx}
\begin{tablenotes}
    \footnotesize
    \item[*] Whether the model has been trained on datasets that include Java code.
    \item[\dag] Whether the model uses instruction tuning to follow task-specific prompts.
\end{tablenotes}
\end{threeparttable}
\caption{Selected Code LLMs for code summarization.}
\label{tab:code_llm_comparison}
\vspace{-12pt}
\end{table}

\textbf{Code Summary Models.}
Code summarization aims to automatically generate concise and meaningful natural-language descriptions of code snippets.
Traditional Seq2Seq models, such as CodeT5~\cite{wang2021codet5} and PLBART~\cite{ahmad2021unified}, employ sequence-to-sequence architectures trained on large-scale paired datasets consisting of source code and human-written descriptions.
In contrast, instruction-tuned models, such as CodeLlama~\cite{roziere2023codellama} and StarCoder~\cite{li2023starcoder}, incorporate additional fine-tuning with structured prompts and task-specific instructions, enabling them to generate more context-aware and adaptive code summaries.
In this paper, we select the four models listed in \autoref{tab:code_llm_comparison} as they are widely used, open-source, and explicitly designed for code summarization tasks. Additionally, since all four models have exposure to Java code~\cite{kocetkovstack, husain2019codesearchnet, lu1codexglue}, they are suited for analyzing decompiled Android applications.

\subsection{Learning-based Malware Analysis}
Traditional machine learning~(ML) based approaches primarily focus on coarse-grained malware analysis, such as family classification and benign-malicious identification~\cite{arp2014drebin, mariconti2016mamadroid}. 
However, fine-grained analysis is essential for deeper malware understanding, moving beyond simple classification~\cite{downing2021deepreflect}.
Recent works explore plugin-based or post-hoc methods, such as explainable AI (XAI) techniques, to extend ML models for interpretable malware analysis. These approaches have been applied to malicious snippet detection~\cite{he2023msdroid, liu2023enhancing}, function identification~\cite{he2023finer}, and behavioral modeling~\cite{he2024dream}, providing insights into why a model detects malware. 

\textbf{LLM-powered Analysis.}
Recent efforts have explored LLM-powered malware detection, primarily operating within the established pipeline of conventional classifiers and leveraging LLMs in two ways: 1)~querying LLMs with original features to generate detection outputs~\cite{malware_decisioncentric}, and 2)~using LLMs to encode text-based semantic representations that enrich traditional feature spaces~\cite{apppoet}.
More advanced approaches leverage GPT-4o-mini's code summarization capabilities~\cite{malware_exploring}, improving malware detection via program slicing techniques and multi-tiered factual checking~\cite{qian2025lamd}.
Despite these advancements, the effectiveness of Code LLMs in fine-grained analysis remains uncharted, largely due to the lack of ground truth. 
Our work goes beyond classification and complements existing studies by systematically benchmarking open-source Code LLMs on both function-level and app-level malware analysis tasks.


\section{Our Evaluation Framework}

We propose a benchmarking framework, named \framework{}, for systematic evaluation of Code LLMs in Android malware analysis.
In this section, we introduce the overview and technical details.

\begin{figure*}[t]
    \centering
    \includegraphics[width=.92\linewidth]{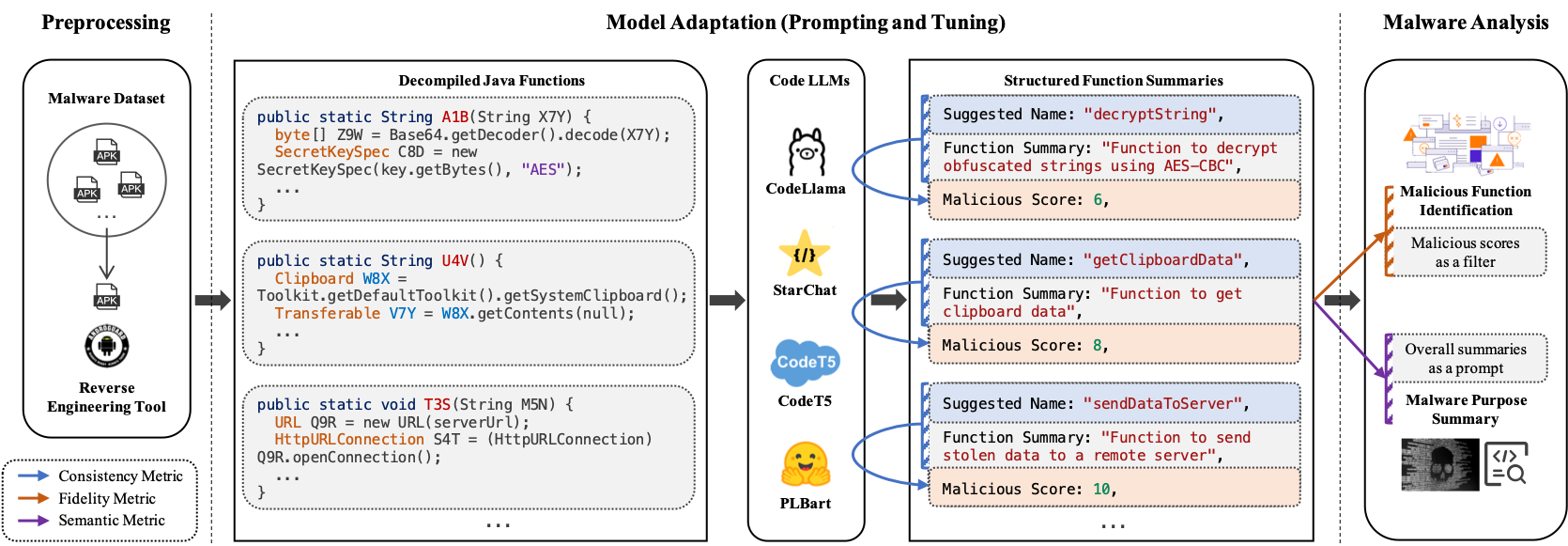}
    \caption{Evaluation pipeline of \framework{}.}
    \label{fig:pipeline}
\end{figure*}

\subsection{Overview}




As illustrated in \autoref{fig:pipeline}, our benchmarking framework is structured into three main stages: dataset preprocessing, model adaptation of Code LLMs, and downstream malware analysis.
\begin{itemize}
    \item \textbf{Dataset Preprocessing.} We first build a representative benchmark dataset by collecting Android malware samples across different malware categories and families. The reverse engineering tool Androguard~\cite{androguard} is used to generate decompiled Java functions for each APK. To ensure function diversity and representativeness, we apply a \textbf{category-wise de-duplication} based on APK size and the number of extracted methods. Since APKs within the same malware category often exhibit only minor variations, this step helps eliminate near-duplicate samples.
    \item \textbf{Model Adaptation of Code LLMs.} Next, we carefully design prompting strategies and tuning procedures for the evaluated code LLMs. Our primary goal is to guide these models to produce \textbf{structured outputs}, specifically consisting of 1)~suggestions of refined \textbf{method names}, 2)~concise and meaningful \textbf{function summaries}, and 3)~\textbf{maliciousness scores} indicating potential harmfulness. Such structured outputs facilitate targeted and interpretable malware analysis.
    \item \textbf{Downstream Malware Analysis Tasks.} We leverage the structured outputs from the LLMs to define two downstream analysis tasks essential for malware characterization: 1)~\textbf{Malicious Function Identification},  where we utilize the maliciousness scores as filters to pinpoint malicious functions within Android apps, enabling analysts to efficiently locate suspicious code segments; and 2)~\textbf{Malware Purpose Summarization}, where we aggregate function-level summaries into comprehensive prompts, supporting the automatic generation of concise malware descriptions detailing their overall malicious objectives and behavior.
\end{itemize}

Within this structure, we design three domain-specific metrics to evaluate model effectiveness in generating structured outputs for downstream tasks.
We introduce details of the model adaptation in \autoref{sec:prompt_tuning} and the three metrics in \autoref{sec:metric}.

\subsection{Prompting and Tuning} \label{sec:prompt_tuning}

To effectively leverage code LLMs for Android malware analysis, we design prompting strategies and tuning mechanisms that guide models to generate the structured outputs.
We adopt two complementary approaches: \textbf{prompt engineering} for instruction-tuned models and \textbf{instruction tuning} for text-to-text models.


\subsubsection{Prompt Engineering}

Prompt engineering involves designing effective input templates to elicit structured responses from models. 
Instruction-tuned models such as StarChat and CodeLlama{}\footnote{We use \texttt{CodeLlama-Instruct} and \texttt{StarChat-Beta}, the instruction-tuned variants of CodeLlama and StarCoder, respectively. These models are optimized for instruct-following code summarization, making them better suited for our task.} are particularly suitable for this approach, as they are optimized for instruction-following and structured generation tasks~\cite{weifinetuned}.
To elicit consistent outputs that include function descriptions, name suggestions, and maliciousness scores, we design prompts that adhere to several key principles:

\begin{itemize} 
\item \textit{Instruction Blocks:} We wrap the main task instruction using special tokens \texttt{[INST]} and \texttt{[/INST]}, following each model’s best practices for instruction prompting. 
\item \textit{Code Delimiters:} Decompiled function code is enclosed between \texttt{[FUNC]} and \texttt{[/FUNC]} tokens to distinguish it from the rest of the prompt and emphasize it as the primary input.
\item \textit{Role Context:} The task is contextualized from the perspective of a \textit{cybersecurity expert} analyzing decompiled Android functions, encouraging the model to reason with a security mindset.
\item \textit{Structured Requirements:} The instruction clearly specifies that the model should return three structured outputs. The expected response is explicitly described in the prompt (see Output Requirements I–III below).
\end{itemize}

\begin{tcolorbox}[title=Prompt I. Structured Function Summarization, 
toprule=.8pt, bottomrule=.8pt, leftrule=.8pt, rightrule=.8pt,
left=2pt, 
right=2pt, 
top=3pt, 
bottom=3pt, 
fonttitle=\small,colback=gray!20, colframe=black, colbacktitle=black, coltitle=white, sharp corners, fontupper=\small, fontlower=\small, before upper=\raggedright, before lower=\raggedright]
[INST]
You are a cybersecurity expert specializing in reverse engineering and malware analysis. Your task is to analyze a decompiled Android function and generate a structured function summary based on the following aspects :

\hspace{5pt}1. \textbf{Function Summary} : $\{summary\_requirement\}$
    
\hspace{5pt}2. \textbf{Suggested Function Name} : $\{name\_requirement\}$
    
\hspace{5pt}3. \textbf{Malicious Score(0-10)} : $\{score\_requirement\}$
[/INST]

[FUNC]
\textbf{$\{decompiled\_code\}$}
[/FUNC]
\end{tcolorbox}

\begin{tcolorbox}[boxsep=0pt, 
toprule=.8pt, bottomrule=.8pt, leftrule=.8pt, rightrule=.8pt,
left=2pt, 
right=2pt, 
top=3pt, 
bottom=3pt, 
fonttitle=\small,colback=white, colframe=black, fontupper=\small, fontlower=\small, before upper=\raggedright, before lower=\raggedright]
Output Requirement I. Summary
\tcblower
<Provide a brief, high-level description of what this function does. Summarize its purpose, key operations, and intent.>
\label{req_summary}
\end{tcolorbox}

\begin{tcolorbox}[boxsep=0pt, 
toprule=.8pt, bottomrule=.8pt, leftrule=.8pt, rightrule=.8pt,
left=2pt, 
right=2pt, 
top=3pt, 
bottom=3pt, 
fonttitle=\small,colback=white, colframe=black, fontupper=\small, fontlower=\small, before upper=\raggedright, before lower=\raggedright]
Output Requirement II. Name
\tcblower
<Suggest a clearer, more descriptive function name that accurately represents its behavior.>
\end{tcolorbox}

\begin{tcolorbox}[boxsep=0pt, 
toprule=.8pt, bottomrule=.8pt, leftrule=.8pt, rightrule=.8pt,
left=2pt, 
right=2pt, 
top=3pt, 
bottom=3pt, 
fonttitle=\small,colback=white, colframe=black, fontupper=\small, fontlower=\small, before upper=\raggedright, before lower=\raggedright]
Output Requirement III. Score
\tcblower
<Rate the function's maliciousness on a scale from 0 to 10, where:
    
\hspace{5pt}0 - Benign : No suspicious activity.
        
\hspace{5pt}1-3 - Potentially Safe but Risky : Performs sensitive actions but could be legitimate.
        
\hspace{5pt}4-6 - Suspicious : Uses permissions or techniques common in malware.
        
\hspace{5pt}7-10 - Highly Malicious : Strong indicators of malware behavior.>
\end{tcolorbox}

\subsubsection{Instruction Tuning}

\begin{figure}[t]
    \centering
    \includegraphics[width=\linewidth]{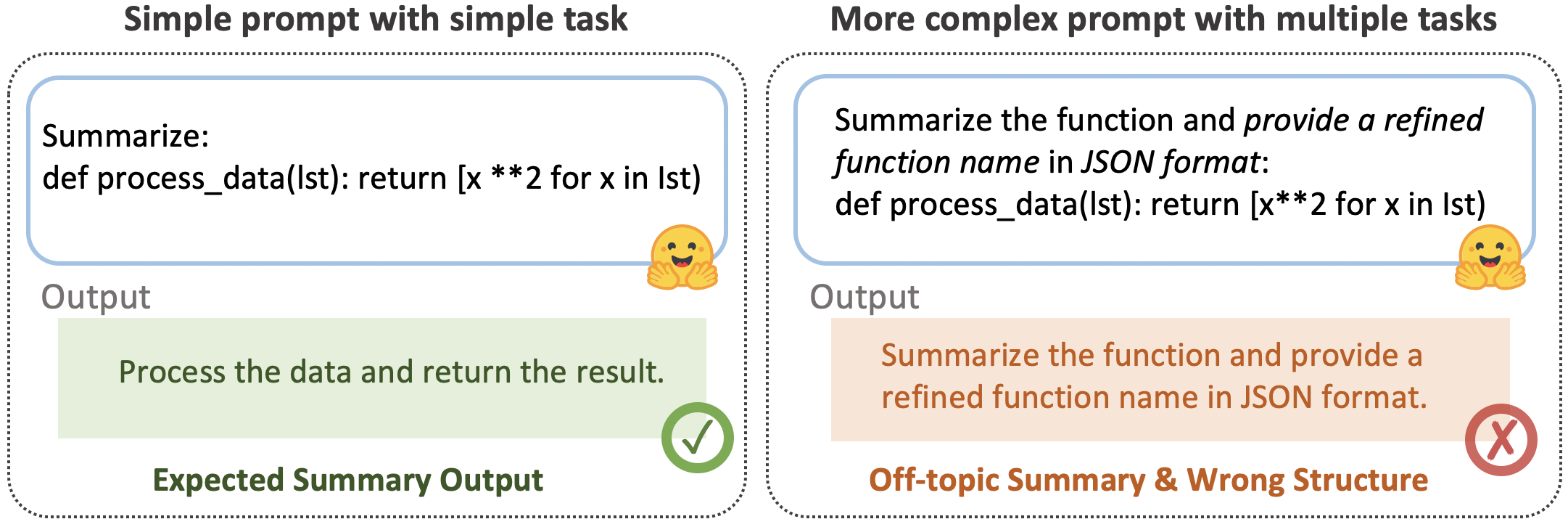}
    \caption[codet5]{Demonstration of the limited capability of CodeT5\footnotemark{} in generating a meaningful output when additional requirements are specified.} 
    \label{fig:codet5_faliure}
\end{figure}

Since models like CodeT5 and PLBART are pretrained for general-purpose code summarization, they inherently lack the ability to generate function names or maliciousness scores. 
For instance, as shown in \autoref{fig:codet5_faliure}, when tasked with generating a function summary alongside a refined function name, CodeT5 would produce off-topic and unstructured responses, failing to generate the expected fields.
To bridge this gap, we apply instruction tuning using task-specific data.

\begin{itemize}
    \item \textit{Function Name Prediction}: We modify the training data by replacing function names with a placeholder (\texttt{unk\_function}). The model is then fine-tuned to predict the actual function name based on the surrounding code. This adaptation allows CodeT5 and PLBART to suggest meaningful function names instead of generic or incomplete descriptions.
    \item \textit{Maliciousness Score Prediction}: Since large-scale ground truth labels for maliciousness scores are unavailable, we introduce a two-step approach. First, we use a tuned model to generate structured summaries and function names. Then, we leverage larger models (e.g., GPT-4 and DeepSeek~\cite{guo2024deepseek}) to infer maliciousness scores based on the generated summaries and function names.
\end{itemize}

This hierarchical approach allows us to enhance the structured output capabilities of CodeT5 and PLBART while leveraging more powerful models for tasks that require higher-level reasoning, such as estimating maliciousness scores.

\footnotetext[\value{footnote}]{\url{https://huggingface.co/Salesforce/codet5-base-multi-sum}}


\subsection{Domain-Specific Metrics} \label{sec:metric}

To rigorously assess the performance of code LLMs in Android malware analysis, we define three domain-specific evaluation metrics: \textbf{consistency}, \textbf{fidelity}, and \textbf{semantic relevance}. These metrics quantitatively measure the effectiveness of structured outputs at different levels—individual function analysis, malware classification, and overall application characterization.

\noindent\textbf{Notations.}
Given an Android application \( \mathcal{A} \) composed of a set of decompiled functions \( \mathcal{F} = \{ f_1, f_2, ..., f_n \} \), our goal is to evaluate the structured outputs generated by a target code LLM~(denoted as \( G \)), including function summary \( S(f) \), refined function name \( N(f) \), and maliciousness score \( M(f) \).
We formally define the \textit{structured output} for function \( f \) as
\(
O(f) = S(f) \oplus N(f) \oplus M(f)
\),
which encapsulates all three elements generated by the LLM~\footnote{The operator \( \oplus \) denotes concatenation and is used consistently throughout the paper.}.
For specific evaluation tasks, we define the \textit{function descriptor} as
\(
D(f) = S(f) \oplus N(f)
\),
which serves as the interpretable textual function representation~(without including its numerical measure) and is particularly relevant in tasks that focus on function-level understanding and classification.


\subsubsection{Consistency-based Metric}

The consistency metric measures the internal stability of the LLM’s structured outputs by checking whether the model’s predictions contradict each other when examined under a self-referential process.
We define two forms of consistency, i.e., \textit{maliciousness consistency} and \textit{name consistency}.

\noindent\textbf{Maliciousness Consistency.}
This metric evaluates whether the maliciousness scores generated by the LLM from raw decompiled code align with those produced when the model is queried with structured descriptors (\textit{function summaries and suggested names}).
Formally, for each function \( f \), we obtain:
\begin{enumerate}[leftmargin=*, label=\alph*)]
    \item \( M_{\text{raw}}(f) \), the original maliciousness score, obtained by directly querying the target model using the decompiled function.
    \item \( M_{\text{des}}(f) \), the descriptor-based maliciousness score, obtained by prompting the same model with \( D(f) \).
\end{enumerate}

\begin{tcolorbox}[title=Prompt II. Descriptor-based Maliciousness Score, 
toprule=.8pt, bottomrule=.8pt, leftrule=.8pt, rightrule=.8pt,
left=2pt, 
right=2pt, 
top=3pt, 
bottom=3pt, 
fonttitle=\small,colback=gray!20, colframe=black, colbacktitle=black, coltitle=white, sharp corners, fontupper=\small, fontlower=\small, before upper=\raggedright, before lower=\raggedright]
\textbf{Task:} Given a function descriptor, $\{score\_requirement\}$ 
\\
\textbf{Input:} A function descriptor: \( \{D_{raw}(f) = S_{raw}(f) \oplus N_{raw}(f) \} \)
\\
\textbf{Output:} A numerical maliciousness score between 0 and 10, where 10 represents highly malicious behavior.
\end{tcolorbox}

To measure consistency, we first normalize the score vectors over all functions in an application~(\(f\in\mathcal{A}\)) into valid probability distributions~(non-negative and summing to 1), obtaining $M^{\prime}_{\text{raw}}$ and $M^{\prime}_{\text{des}}$.
Then we compute the distributional divergence using Jensen-Shannon Divergence (JSD):
\begin{equation}
\text{JSD}(M^{\prime}_{\text{raw}}, M^{\prime}_{\text{des}}) = \frac{1}{2} D_{\text{KL}}(M^{\prime}_{\text{raw}} || M_{\text{avg}}) + \frac{1}{2} D_{\text{KL}}(M^{\prime}_{\text{des}} || M_{\text{avg}}) \,,
\end{equation}
where \( M_{\text{avg}} \) is the average distribution and \( D_{\text{KL}}(P || Q) \) is the Kullback-Leibler divergence:
\begin{equation}
M_{\text{avg}} = \frac{1}{2}(M^{\prime}_{\text{raw}} + M^{\prime}_{\text{des}}) \,,
D_{\text{KL}}(P || Q) = \sum_{i} P(i) \log \frac{P(i)}{Q(i)} \,.
\end{equation}
Finally, we normalize the JSD to $(0,1)$ and define: 
\begin{equation}
\text{MCS} = 1 - \frac{\text{JSD}(M^{\prime}_{\text{raw}}, M^{\prime}_{\text{des}})}{\log 2} \,.
\end{equation}

A higher maliciousness consistency score~(MCS) indicates higher consistency, meaning that the structured outputs retain the function’s security-relevant information. 

\noindent\textbf{Name Consistency.}
This metric assesses whether the suggested function name remains stable when the LLM is prompted with its own function summary.
For each function \( f \), the LLM generates:
\begin{enumerate}[leftmargin=*, label=\alph*)]
    \item \( N_{\text{raw}}(f) \), the initial function name suggested as part of the structured output \( O(f) \).
    \item \( N_{\text{reg}}(f) \), a new function name generated when the LLM is re-prompted with its own function summary \( S(f) \).
\end{enumerate}

\begin{tcolorbox}[title=Prompt III. Re-generated Function Name, 
toprule=.8pt, bottomrule=.8pt, leftrule=.8pt, rightrule=.8pt,
left=2pt, 
right=2pt, 
top=3pt, 
bottom=3pt, 
fonttitle=\small,colback=gray!20, colframe=black, colbacktitle=black, coltitle=white, sharp corners, fontupper=\small, fontlower=\small, before upper=\raggedright, before lower=\raggedright]
\textbf{Task:} Given a function summary, $\{name\_requirement\}$ 
\\
\textbf{Input:} A function summary: \( \{ S_{raw}(f) \} \) 
\\
\textbf{Output:} A concise, descriptive function name.
\end{tcolorbox}

To quantify name consistency, we compute the normalized edit distance between the original and revised function names:
\begin{equation}
\text{NCS} = 1 - \frac{\text{EditDistance}(N_{\text{raw}}, N_{\text{reg}})}{\max(|N_{\text{raw}}|, |N_{\text{reg}}|)} \,,
\end{equation}
where $|n|$ represents the length of $n$, and EditDistance is the Levenshtein distance, which counts the minimum number of character-level insertions, deletions, or substitutions required to transform \( N_{\text{raw}} \) into \( N_{\text{reg}} \). 
The result is normalized to \( (0,1) \) by the length of the longer function name to ensure comparability across different naming conventions.

A higher name consistency score~(NCS) indicates greater stability in function name generation, suggesting that the model consistently associates summaries with the same function identity. 


\subsubsection{Fidelity-based Metric}

The fidelity metric assesses the degree to which function-level structured outputs contribute to \textit{malicious function identification}. Inspired by explainable AI (XAI) evaluation techniques~\cite{yang2019evaluating}, we define fidelity in terms of the impact of function removals on malware classification performance.

Given a malware classifier \( C \), which takes the function descriptor as input and predicts a malware category \( \hat{y} \), we measure classification confidence before and after removing the \textit{top-$k$ most malicious} function summaries.
Formally, let:
\begin{equation}
p_{\text{full}} = C(D(f_1) \oplus D(f_2) \oplus ... \oplus D(f_n))
\end{equation}
be the malware classification probability for an application before removal. After removing the top-$k$ most malicious function features ranked by maliciousness, the new classification probability is:
\begin{equation}
p_{\text{red}(k)} = C\left(\textstyle\bigoplus_{f \notin \mathcal{F}_k} D(f)\right) \,,
\end{equation}
where \( \mathcal{F}_k = \{ f_i \in \mathcal{F} \mid M(f_i) \text{ is among the top } k \} \) is the set of \textit{k most malicious functions}. 
The maliciousness-based fidelity score~(MFS) is then computed as the \textit{relative drop in confidence}:
\begin{equation}
\text{MFS}_{(k)} = \frac{p_{\text{full}}[\hat{y}] - p_{\text{red}(k)}[\hat{y}]}{p_{\text{full}}[\hat{y}]} \,.
\end{equation} 

A higher fidelity score indicates that structured outputs effectively encode function-level characteristics for malware classification, as the maliciousness-based descriptor removal leads to a significant drop in the classifier’s confidence for the predicted class.

\subsubsection{Semantic-based Metric}

The semantic metric evaluates how function outputs contribute to accurate application-level \textit{malware purpose descriptions}. Adopting approaches from automatic machine translation evaluation~\cite{mathur2020tangled}, we measure the similarity between LLM-generated malware descriptions and reference descriptions. 

Given a set of top-$v$ malicious function outputs, where \( v \) varies based on the target LLM’s context window, we firstly generate 
\begin{equation}
A_{\text{LLM}} = G_{app}(O(f_1) \oplus O(f_2) \oplus ... \oplus O(f_v)) \,,
\end{equation}
where \( G_{app} \) is the target code LLM prompted to generate a high-level malware description.
This approach mimics \textit{context slicing}, but leverages LLM outputs~(i.e., maliciousness scores) instead of heuristic-based methods such as sensitive API filtering, which can often be incomplete or overlook critical behaviors.
Specifically, the prompt for \( G_{app} \) is defined as follows.

\begin{tcolorbox}[title=Prompt IV. Application Purpose Description, 
toprule=.8pt, bottomrule=.8pt, leftrule=.8pt, rightrule=.8pt,
left=2pt, 
right=2pt, 
top=3pt, 
bottom=3pt, 
fonttitle=\small,colback=gray!20, colframe=black, colbacktitle=black, coltitle=white, sharp corners, fontupper=\small, fontlower=\small, before upper=\raggedright, before lower=\raggedright]
\textbf{Task:} Given the structured function-level analyses, generate a concise and comprehensive description of the overall application's purpose.  
\\
\textbf{Input:} A set of top-\( v \) malicious functions:  
\{ Function Summary \( S(f) \), Refined Function Name \( N(f) \), Maliciousness Score \( M(f) \) \}  
\\
\textbf{Output:} An application purpose description summarizing the app’s behavior and potential security risks.
\end{tcolorbox}

We compare \( A_{\text{LLM}} \) against the reference malware description \( A_{\text{GT}} \) using three widely used text similarity metrics.
BLEU~\cite{papineni2002bleu} measures n-gram precision between \( A_{\text{LLM}} \) and \( A_{\text{GT}} \):
\begin{equation}
\text{BLEU}(A_{\text{LLM}}, A_{\text{GT}}) = \exp \left( \sum_{n=1}^{N} w_n \log p_n \right)
\end{equation}
where \( p_n \) is n-gram precision and \( w_n \) are weighting factors.
Additionally, we report METEOR~\cite{banerjee2005meteor}, which incorporates synonym matching and recall, and ROUGE-L~\cite{lin2004rouge}, which evaluates the longest common subsequence overlap.


A higher BLEU, METEOR, or ROUGE-L score indicates stronger alignment between the LLM-generated description and the reference description, validating the semantic relevance of function outputs in capturing malware behavior.



\begin{table*}[t]
\begin{threeparttable}
\begin{smalltabularx}{\linewidth}{C|CC|CCC|CCC}
\toprule
\multicolumn{1}{c}{\multirow{2}{*}{}} & \multicolumn{2}{|c}{\textbf{Consistency}} & \multicolumn{3}{|c}{\textbf{Fidelity}} & \multicolumn{3}{|c}{\textbf{Semantic Relevance}} \\ 
\cmidrule(lr){2-3}\cmidrule(lr){4-6}\cmidrule(l){7-9}
\multicolumn{1}{c}{}    & \multicolumn{1}{|c}{MCS} & \multicolumn{1}{c}{NCS} & \multicolumn{1}{|c}{$\text{MFS}_{(2)}$} & \multicolumn{1}{c}{$\text{MFS}_{(5)}$} & \multicolumn{1}{c}{$\text{MFS}_{(8)}$} & \multicolumn{1}{|c}{BLEU} & \multicolumn{1}{c}{METEOR} & \multicolumn{1}{c}{ROUGE-L} \\ 
\midrule
\textbf{CodeT5}   & N/A              & $0.233\pm0.04$         & $0.332\pm0.30$          & $0.125\pm0.22$          & $0.396\pm0.32$          & $0.059\pm0.04$ & $0.083\pm0.04$ & $0.186\pm0.04$ \\
\textbf{PLBART}   & N/A              & $0.499\pm0.05$         & $0.033\pm0.14$          & $0.031\pm0.11$          & $0.065\pm0.14$          & $0.137\pm0.03$ & $0.185\pm0.05$ & $0.228\pm0.04$ \\ 
\textbf{CodeLlama} & $0.381\pm0.03$    & \cellcolor[HTML]{e0e0e0}$0.628\pm0.04$         & \cellcolor[HTML]{e0e0e0}$0.158\pm0.27$          & $0.159\pm0.27$          & $0.113\pm0.25$          & $0.175\pm0.05$ & $0.247\pm0.08$ & $0.271\pm0.06$ \\
\textbf{StarChat} & \cellcolor[HTML]{e0e0e0}$0.813\pm0.02$    & $0.575\pm0.02$         & $0.111\pm0.20$          & \cellcolor[HTML]{e0e0e0}$0.254\pm0.30$          & \cellcolor[HTML]{e0e0e0}$0.275\pm0.33$          & \cellcolor[HTML]{e0e0e0}$0.176\pm0.05$ & \cellcolor[HTML]{e0e0e0}$0.273\pm0.09$ & \cellcolor[HTML]{e0e0e0}$0.272\pm0.06$ \\ 
\midrule
\textbf{CodeLlama+} & 0.357 \percplus{-6.30}   & 0.677 \percplus{7.80}    & 0.485 \perc{206.96}  & 0.451 \perc{183.65}  & 0.440 \perc{289.38}   & 0.171 \percplus{-2.29} & 0.219 \percplus{-11.34} & 0.270 \percplus{-0.37} \\
\textbf{StarChat+}  & 0.828 \percplus{1.85}   & 0.582 \percplus{1.22}    & 0.298 \perc{168.47}  & 0.351 \percplus{38.19}  & 0.726 \perc{164.00}   & 0.172 \percplus{-2.27} & 0.246 \percplus{-9.89} & 0.274 \percplus{0.74} \\ 
\bottomrule 
\end{smalltabularx}
\begin{tablenotes}
    \footnotesize
    \item[*] Rows 1–4 correspond to RQ1, evaluating the performance of all four models on decompiled code. Results are reported as mean $\pm$ standard deviation.
    \item[*] Rows 5–6 correspond to RQ2, assessing the impact of function renaming by replacing original names with LLM-suggested ones. CodeT5 and PLBART are excluded due to limited name generation capability—they often replicate names from the input code. Results are reported as mean values with relative improvement ratios.
\end{tablenotes}
\end{threeparttable}
\caption{Benchmarking results.}
\label{tab:benchmark_results}
\vspace{-8pt}
\end{table*}

\section{Benchmarking Results}

We conduct experiments guided by two key research questions:
\begin{itemize} 
\item \textbf{RQ1:} How well do Code LLMs understand decompiled code for malware analysis tasks? 
\item \textbf{RQ2:} How does function renaming affect their performance~(i.e., can models self-repair based on their own suggested names)? \end{itemize}


\subsection{Experimental Setup} 


\noindent\textbf{Dataset Selection.}
We use the LAMD~\cite{qian2025lamd} dataset\footnote{LAMD dataset collected in the 2020s: \url{https://zenodo.org/records/14884736}}, which provides Android malware samples with high-quality GPT-4-generated ground-truth application purpose summaries, making it well-suited for evaluating semantic relevance.
To reduce redundancy, we perform de-duplication by filtering out near-identical APKs within each malware category, resulting in 118 APKs across 6 categories~(Adware, Backdoor, PUA, Riskware, Scareware, Trojan) and 13 families. 
All APKs are decompiled using Androguard, resulting in a total of $7,542,799$ decompiled functions across the dataset. 

\noindent\textbf{Implementation Details.}
For all selected models in \autoref{tab:code_llm_comparison}, we use their official implementations from the Hugging Face Hub~\footnote{\url{https://huggingface.co/}\{meta-llama/CodeLlama-7b-Instruct-hf, HuggingFaceH4/starchat-beta, Salesforce/codet5-base-multi-sum, uclanlp/plbart-base\}}. 
For models not originally instruction-tuned (i.e., CodeT5 and PLBART), we perform additional tuning on Java functions from their pretraining datasets.
Each model is fine-tuned for 3 epochs, which we find sufficient to produce the structured outputs.
The maliciousness score prediction of these two models is assisted by a locally deployed DeepSeek-R1-Distill-Llama-70B.
In the fidelity evaluation, we use \texttt{LightGBM}~\cite{ke2017lightgbm} as the malware category classifier $C$, ensuring high reliability with an accuracy above $0.95$. 
To assess the effect of removing suspicious code, we experiment with top-$k$ values of $2$, $5$, and $8$.
In the semantic relevance evaluation, to ensure stylistic consistency, we prompt all models to begin their outputs with the phrase: \textit{``This application appears to...''}, matching the format used in the ground truth.
We set top-$v$ based on model context limits: $4$K tokens for CodeLlama, $8$K for StarChat, and $1$K for CodeT5 and PLBART.
For BLEU-based evaluation, we use 2-gram precision, which is more appropriate for evaluating short summaries.


The generated outputs are made open-source~\footnote{Our \framework{} dataset: \url{https://zenodo.org/records/15155917}}.
Our overall experimental results are summarized in \autoref{tab:benchmark_results}. 
In the following sections, we provide a detailed analysis of each research question: RQ1 in \autoref{evaluation_rq1} and RQ2 in \autoref{evaluation_rq2}.

\subsection{RQ1 - Decompiled Code} \label{evaluation_rq1}

This experiment investigates how well Code LLMs interpret decompiled Android code for malware analysis. We evaluate their ability to generate the structured outputs and analyze their effectiveness using the three domain-specific metrics.

For maliciousness consistency, only CodeLlama and StarChat are evaluated, as CodeT5 and PLBART rely on external models for score generation. 
Among the two, StarChat achieves a notably higher score (over twice that of CodeLlama), suggesting a better understanding of high-level malware semantics.  
For name consistency, CodeLlama performs best, with StarChat following closely. 
Both models outperform the Seq2Seq baselines, reinforcing the observation that instruction-tuned, GPT-style models exhibit greater stability in structured output generation. 
Among the Seq2Seq models, PLBART outperforms CodeT5 by approximately $114\%$, likely due to its improved alignment between code and natural language.

In downstream evaluations, we observe clear performance differences across models in both fidelity and semantic relevance. 
StarChat consistently outperforms the others, demonstrating a stronger ability to assign meaningful maliciousness scores and produce high-level malware descriptions—results that align with its superior consistency metrics.
This superior performance is likely driven by its larger model size and the StarCoder architecture, which emphasizes multilingual understanding and instruction-following. 
CodeLlama performs competitively, especially excelling in top-2 function removal and producing stylistically aligned summaries, suggesting it effectively captures the most critical functions but is less robust than StarChat when evaluating broader function sets.

Among the Seq2Seq models, PLBART shows better performance in semantic relevance, benefiting from its BART-based architecture, which favors fluent and coherent natural language generation. 
However, PLBART notably underperforms CodeT5 in fidelity, while both models' maliciousness scores are generated externally by the same larger models.
This difference arises because CodeT5’s summaries, though less fluent, contain more descriptions that better highlight critical code features, allowing the external scoring model to produce more discriminative maliciousness scores.
Nevertheless, both Seq2Seq models exhibit limited stylistic control. 
For instance, even when explicitly prompted to begin with \textit{``This application appears to...''}, they frequently prepend generic phrases like \textit{``This function...''}, revealing limited control over stylistic constraints.

These findings highlight a fundamental \textbf{tradeoff between linguistic fluency and semantic precision} in LLM-generated outputs.
Our results also reinforce the \textbf{superiority of instruction-tuned GPT-style models~(especially StarChat)} for generating both accurate and interpretable outputs in fine-grained malware analysis tasks.

\subsection{RQ2 - Function Naming} \label{evaluation_rq2}

This experiment investigates whether replacing original decompiled function names with LLM-suggested names affects model performance. The goal is to understand whether LLMs can improve their own reasoning and potentially providing more meaningful input for subsequent predictions.

We compare each model’s outputs before and after replacing function names, modifying only those names that differ from the originals, while keeping all other code aspects unchanged.
CodeT5 and PLBART are excluded from RQ2 due to their limited function name refinement capability. 
These models often fail to produce meaningful or distinct name suggestions. 
For instance, $61.75\%$ of PLBART’s generated names are exact copies of the names found in the decompiled code. Even among the remaining cases, many suggestions are only trivially modified~(e.g., renaming \texttt{set\_b} to \texttt{set\_a}), which lacks usefulness in evaluating the impact of renaming. 

\begin{figure}[t]
  \centering
  \begin{subfigure}[b]{0.49\linewidth}
    \centering
    \includegraphics[width=\linewidth]{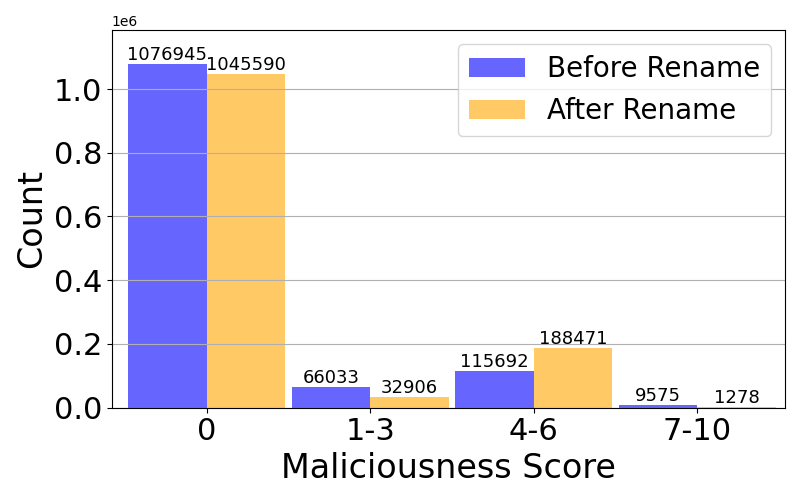}
    \caption{CodeLlama}
    \vspace{-2pt}
    \label{fig:fig1}
  \end{subfigure}\hfill
  \begin{subfigure}[b]{0.49\linewidth}
    \centering
    \includegraphics[width=\linewidth]{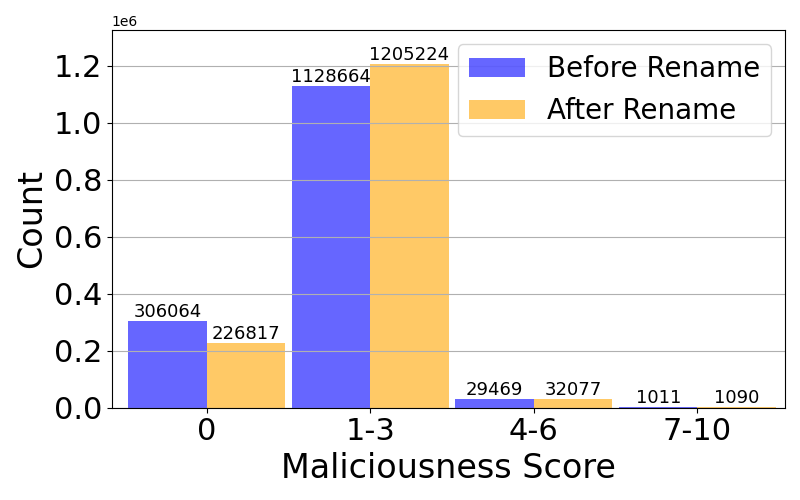}
    \caption{StarChat}
    \vspace{-2pt}
    \label{fig:fig2}
  \end{subfigure}
  \caption{Maliciousness score distributions before and after function renaming. For both models, refined function names lead to more scores concentrated in the middle range.}
  \label{fig:combined}
\end{figure}

Results show that replacing original function names with LLM-suggested names notably benefits fidelity, with both CodeLlama+ and StarChat+ showing substantial improvements. This suggests that improved naming significantly helps the models better prioritize and identify critical malicious functions. 
Consistency also improves moderately, indicating enhanced stability in model predictions when meaningful names are used. 
However, semantic relevance slightly decreases after renaming, likely because renaming leads to a convergence of maliciousness scores around the mid-range~(\autoref{fig:combined}), reducing the distinctiveness of highly ranked functions when aggregated into malware descriptions. 

Overall, these findings indicate that LLM-based function renaming effectively \textbf{enhances function-wise consistency and fidelity metrics}, but may \textbf{require careful handling to avoid diluting high-level semantic clarity}.
To mitigate this issue, future improvements could focus on calibrating the scores to better reflect model confidence and explicitly encoding more robust knowledge about malware semantics. 

\section{Discussion}



While \framework{} enables structured evaluation of Code LLMs in malware analysis, it also exposes a \textit{fundamental challenge}: the scarcity of reliable ground truth at the function and behavior levels. Our function-level evaluation relies on counterfactual fidelity-based methods, while APK-level summarization adopts techniques from LLM-driven malware detection~\cite{qian2025lamd}, which leverage program slicing and prompt large models like GPT-4. Though practical, these surrogate approaches can introduce noise and bias~\cite{she2023pitfalls}, which underscores the pressing need for high-quality, fine-grained ground truth malware datasets to advance trustworthy evaluation.

\framework{} opens the door to \textit{broader research directions}. 
For example, it can support studies on malware concept drift~\cite{pendlebury2019tesseract}, enabling evaluation of whether Code LLM-based analyses generalize to evolving threats.
Beyond Code LLMs and Android malware, the framework is adaptable to assess a wide range of approaches, as long as they target core sub-tasks including function summarization, naming, and maliciousness estimation.
Beyond benchmarking, \framework{} also supports \textit{practical applications}: it can guide the selection, pretraining, or fine-tuning of Code LLMs specifically for malware tasks~\cite{al2023extending, yang2024tapi, shao2025explanation}. Its structured outputs, particularly maliciousness scores, can be used to prioritize suspicious functions, improving the precision of traditional malware classifiers~\cite{arp2014drebin, he2023msdroid}.
\section{Conclusion}

This paper introduces \framework{}, a benchmarking framework for systematically evaluating the effectiveness of open-source Code LLMs in Android malware analysis. We define a structured output format aligned with two key analysis tasks: malicious function identification and malware purpose summarization.
To address the lack of fine-grained ground truth, we propose three domain-specific evaluation metrics, enabling rigorous assessment of LLM-generated outputs.
Our benchmarking results reveal both the potential and current limitations of Code LLMs.
\framework{} provides a foundation for future work to select and adapt Code LLMs for malware analysis, improving their effectiveness in downstream tasks such as family classification and behavioral explanation.




\balance
\bibliographystyle{ACM-Reference-Format}
\bibliography{llma}

\end{document}